# The Nature of Plerions Surrounding Soft Gamma-Ray Repeaters

Alice K. Harding

*Laboratory for High Energy Astrophysics, Code 665,
NASA/Goddard Space Flight Center, Greenbelt, MD 20771, USA*

Compact steady sources of X-ray emission have been detected at the positions of at least two soft gamma-ray repeaters (SGRs). These sources have been interpreted as synchrotron nebulae powered by the neutron star that is causing the bursts. We explore a plerion model for the sources surrounding SGRs where the steady observed emission is powered by the SGR bursts rather than by the spin-down of a pulsar. In this case there is no limit on the neutron star magnetic field. We find that the synchrotron lifetime of the particles injected into the plerion around SGR1806-20 is long enough to smear out nebular emission from individual bursts. Transient nebular emission would therefore not be detected following an SGR burst. The combined radio emission from multiple burst injections is expected to have a steeper spectrum than that of a typical plerion.

## INTRODUCTION

Some 20% of the supernova remnants in our galaxy are plerions, or filled synchrotron nebulae, many of them surrounding young pulsars that supply the power to the nebulae through injection of relativistic particles. Plerion-like sources have been detected at the positions at least two of the known soft gamma-ray repeaters, SGR1806-20 (1) (2) and SGR0525-66 (3) in the SNR N49 (4). The third known SGR, SGR1900+14, is near a ROSAT source that is possibly a plerion (5). These associations have firmly established that this class of $\gamma$-ray bursts is linked to neutron stars. It has been suggested that the plerions around SGRs are powered by the quiescent, steady spin-down luminosity of a young pulsar that is also the source of the bursts. However, the fact that the measured radio spectral index ($\alpha = -0.6$) of SGR1806-20 (6) is much steeper than the typical plerion spectrum having indices in the range $\alpha = 0$ to -0.3, makes the pulsar-powered plerion model less than satisfactory. This paper will explore an alternative hypothesis, that the SGR plerions are powered primarily by particle injection from the SGR bursts (7). Drawing on analogies to the pulsar wind model of plerions, we investigate the characteristics expected of nebular synchrotron emission resulting from episodic injections of particles from the SGR bursts, including the suggestion (7) that nebular emission due to individual bursts may be detectable. This model will be applied to SGR 1806-20, where it is energetically plausible that the steady X-ray source observed by ASCA (2) is powered by SGR burst particle injection.

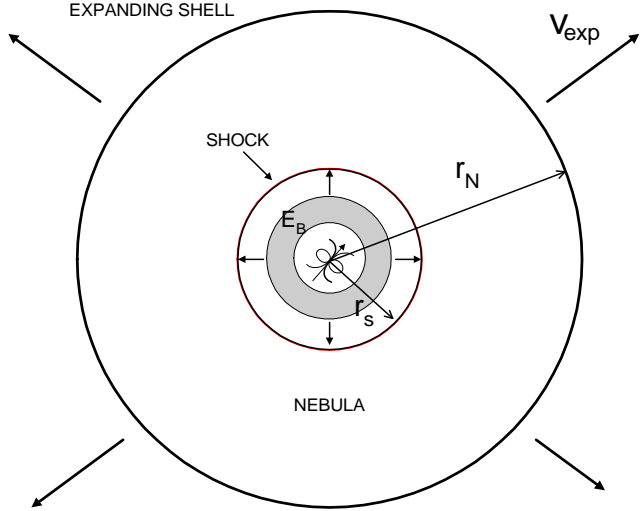

**FIG. 1.** Schematic illustration of the wind model of a plerion.

## WIND MODEL OF PLERIONS

Suppose the source that powers an SGR plerion injects a total *time-averaged* luminosity $\langle L_T \rangle$ in particles and magnetic field in the form of a relativistic MHD wind. Because this wind is confined by the supernova shell which expands with velocity $v \ll c$, there will be a reverse shock at a distance $r_s$ (see Fig 1) from the source where the wind ram pressure balances the nebular pressure (8),

$$\frac{\langle L_T \rangle}{4\pi r_s^2 c} = \frac{\int L_T dt}{\frac{4}{3}\pi r_N^3} \qquad (1)$$

Using the measured radius of the SGR 1806-20 radio nebula at 0.3 GHz (6), $r_N = 5' d_{10}$, where $d_{10} \equiv d/10$ kpc is the source distance, and assuming an age $\tau_4 \equiv \tau/10^4$ yr,

$$r_s \simeq \left(\frac{r_N^3}{3c\tau}\right)^{1/2} = 4 \times 10^{18} \text{ cm } d_{10}^{3/2} \tau_4^{-1/2}. \qquad (2)$$

The position of the shock in this case is not expected to move in response to injection from individual bursts because the propagation time from the burst source to $r_s$ is $\Delta \tau_s \simeq r_s/c \sim 5$ yr. Note also that $r_s \simeq 0.1\, r_N$ in this source is much larger that the shock radius in the Crab nebula, $r_s = 3 \times 10^{17}$ cm, due to its larger size.

## NEBULAR EMISSION FOLLOWING AN SGR BURST?

If an individual SGR burst causes a transient injection of particles and magnetic field which moves out into the nebula, will the synchrotron radiation from that burst be detectable in X-rays above the steady source luminosity? A burst of total energy $E_B$ and duration $\Delta t_B$ will cause a transient wave of particles and field in the nebular wind flow (see Fig 1). Assuming that the source produces $\gamma$-rays with an efficiency, $\epsilon \equiv 10^{-2}\epsilon_{-2}$, then the total luminosity of the burst outflow is $L_B = 10^{40}\,\mathrm{erg\,s^{-1}} L_{40}\,\epsilon^{-1}\,d_{10}^2$, where $L_{40}$ is the burst $\gamma$-ray luminosity. Although the MHD wind in the Crab nebula is out of equipartition by several orders of magnitude upstream of the shock, the ratio of magnetic to particle energy is $\sim .003$, equipartition is achieved shortly downstream of the shock where the bulk of the synchrotron radiation is emitted (9). The magnetic field in the burst outflow, assuming equipartition downstream of the shock and isotropic outflow, is

$$B_{\mathrm{eq}} = \left(\frac{L_B}{c}\right)^{1/2} \frac{1}{r_s} = 1.5 \times 10^{-3}\,\mathrm{G}\ \ L_{40}^{1/2}\,d_{10}^{-1/2}\,\tau_4^{1/2}\,\epsilon_{-2}^{-1/2}. \qquad (3)$$

The synchrotron lifetime of the particles emitting radiation in the outburst at frequency $\nu_{\mathrm{keV}} \equiv \nu/1$ keV will then be

$$\tau_s = 1.2\,\mathrm{yr}\ \nu_{\mathrm{keV}}^{-1/2}\,L_{40}^{-3/4}\,d_{10}^{3/4}\,\tau_4^{-3/4}\,\epsilon_{-2}^{3/4}. \qquad (4)$$

The observed luminosities (10) of bursts from SGR 1806-20 are in the range $L_{40} = 0.8 - 50$, giving $\tau_s \sim 0.07 - 1.4$ yr, with the brightest bursts producing the shortest bursts of nebular radiation. However, the average time between bursts is (11) $\Delta T \simeq 10^5$ s $\ll \tau_s$, with no apparent correlation between burst luminosity and time to the next burst. Therefore, the emission from the bursts will be blended together, producing a continuous flux. The very brightest bursts might produce slight fluctuations in the nebular emission, but only if the next bright burst follows by more than $\tau_s \sim 0.07$ yr and the luminosity will only be a factor of 2-3 above the steady observed X-ray source luminosity of $L_X \simeq 1 \times 10^{35}$ erg s$^{-1}$ (2). Even then, the long propagation time, $\Delta \tau_s \sim 5$ yr, for the particles to reach the shock where their pitch angles randomize and they can emit radiation, will make it impossible to identify an increase in nebular emission with an individual burst. Tavani (7) found that nebular emission may be detectable following an SGR burst if the particle injection is beamed. However, he assumed a smaller shock radius $r_s \sim 10^{16} - 10^{17}$ cm, which gives a much larger $B_{\mathrm{eq}}$ and a much shorter $\tau_s$. Allowing for burst particles to be beamed into a solid angle $\Omega$ will not change the estimate of the $B_{\rm eq}$ in Eqn (3) or $\tau_s$ in Eqn (4) because the burst energy and volume of the flow both decrease as $\Omega$, provided that the particle flow and $\gamma$-ray emission have the same solid angle, giving the same energy density in the outflow.

## SGR POWERED PLERIONS

If the emissions from episodic burst particle injections blend into one another due to the long synchrotron radiation timescales, it raises the question of whether the SGR bursts alone are energetically capable of powering the X-ray nebula. To explore this, we estimate the ratio of observed time-averaged $\gamma$-ray luminosity, $\langle L_B \rangle = \langle E_B \rangle / \langle \Delta T \rangle$ to total luminosity required to power the nebula, $L_T = L_X/\eta = 1 \times 10^{37}\,{\rm erg\,s}^{-1}\,\eta_{-2}^{-1}\,d_{10}^2$, where $\eta_{-2} = \eta/10^{-2}$ is the efficiency of nebular emission. This ratio is,

$$\epsilon_\gamma \approx \frac{\langle L_B \rangle}{\langle L_T \rangle} = \frac{\langle E_B \rangle}{\langle \Delta T \rangle} \frac{\eta}{L_X} \sim 1 \times 10^{-3}\,\eta_{-2} \left(\frac{\Delta T}{10^5\,{\rm s}}\right)^{-1} \left(\frac{E_B}{10^{39}\,{\rm erg}}\right), \qquad (5)$$

which is not unreasonable.

The synchrotron spectrum from a nebula powered by SGR burst injections will be that of the combined emission from many individual bursts. If, as assumed, the particles injected from each burst propagate outward at the bulk velocity of the wind flow, each carrying a "frozen-in" equipartition field, $B_{\rm eq} \propto L_B^{1/2}$ (cf. Eqn (3), the synchrotron spectrum due to each burst will have a break frequency (see Fig 2),

$$\nu_B = 12.6\,{\rm Hz}\,B_{\rm eq}^{-3}\,\tau_4^{-2} = 3.8 \times 10^9\,{\rm Hz}\,L_{40}^{-3/2}\,d_{10}^{3/2}\,\epsilon_{-2}^{3/2}\,\tau_4^{-7/2}, \qquad (6)$$

where $\tau_s = \tau$, above which synchrotron losses become important. The burst luminosity range of SGR 1806-20 corresponds to a range of $\nu_B = 10$ MHz - 5 GHz. The spectrum of the combined burst emissions may be obtained by integrating the synchrotron emissivity, $P(\nu, \gamma)$, over the the burst luminosity function $N_B(L_B) \propto L_B^{-\delta}$ and the steady-state particle spectrum, $N_e(\gamma, L_B)$. We find a synchrotron spectrum,

$$F_{\rm syn}(\nu) \propto \begin{cases} \nu^{-\Gamma_1} = \nu^{-(\alpha-1)/2}, & \nu < \nu_B(L_{\max}) \\ \nu^{-\Gamma_2} = \nu^{-(2\alpha+3-2\delta)/3}, & \nu_B(L_{\max}) < \nu < \nu_B(L_{\min}) \\ \nu^{-\Gamma_3} = \nu^{-\alpha/2}, & \nu > \nu_B(L_{\min}) \end{cases}$$

where $\alpha$ is the electron injection index. In the region $\nu_B(L_{\max}) < \nu < \nu_B(L_{\min})$ between the break frequencies of the brightest and weakest bursts, only emission from bursts below a luminosity $L_B \propto \nu^{-2/3}$ contribute significantly to the spectrum, producing a steepening of the power law from the usual index $(\alpha - 1)/2$ expected from a particle distribution of index $\alpha$, as illustrated in Fig. 2.

If we match the index of the radio spectrum of SGR 1806-20, -0.6, to that of region 2, $-\Gamma_2$, then the inferred injection index of the particle spectrum, assuming it is similar for all bursts, is $\alpha = 1.4$, adopting the luminosity function with $\delta = 2$ derived by Laros (11). This then gives a radio spectral index for $\nu < 10$ MHz of $\Gamma_1 = 0.2$, in the characteristic range for observed plerions. Above $\nu = 5$ GHz, the spectrum will steepen to a index of $\Gamma_3 = 0.7$ when synchrotron losses affect emission from all the bursts. While the break at 5 GHz is well determined by the turnover $L_{\min} \sim 0.8\, L_{40}$ of the observed burst luminosity function (11), the detection of a break at $\nu \leq 10$ MHz would measure the maximum luminosity $L_{\max}$ of the bursts.

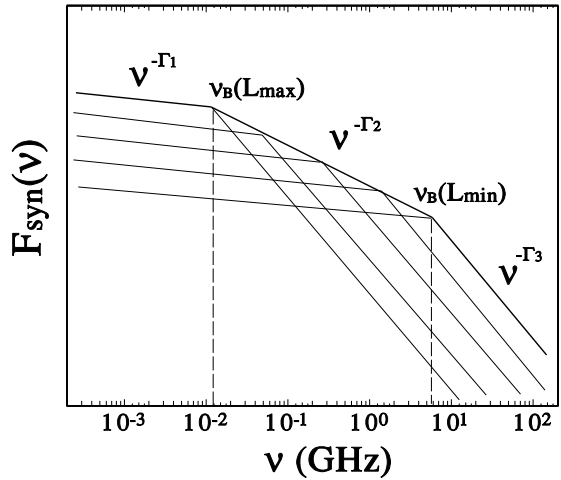

**FIG. 2.** Expected synchrotron spectrum of a plerion powered by bursts from a soft-$\gamma$-ray repeater.

As is observed for the Crab nebula, above the frequency where synchrotron losses steepen the spectrum, the size of the nebula begins to shrink as the distance which the radiating particles travel before losing their energy decreases. This model may therefore be tested through measurements of the size of the radio nebula as a function of frequency.